\documentstyle[12pt]{article}
\begin{document}
\title{LOW DIMENSIONAL ELECTRONS}
\author{B.G. Sidharth$^*$\\
B.M. Birla Science Centre, Hyderabad 500 063 (India)}
\date{}
\maketitle
\footnotetext{$^*$E-mail:birlasc@hd1.vsnl.net.in}
\begin{abstract}
Motivated by recent work on nano tubes indicating one dimensional quantum 
effects and studies of two dimensional electron gas, we consider a recent 
model of Fermions as Kerr-Newman type black holes with quantum mechanical
effects. Such an anyonic fractionally charged quasi-electron picture is
deduced and identified with quarks.
\end{abstract}
Very recently experiments with nano tubes reveal one dimensional Quantum
behaviour\cite{r1,r2}. On the other hand the study of the fractional
quantum Hall effect leads to the study of a two dimensional electron
gas, which can be realised for example in $G_aAs-Al_xGa_{1-x}As$
semiconductor heterojunctions(\cite{r3,r4} and references therein).
In these low dimensions, we have to consider anyonic statistics with
interesting applications\cite{r5}.\\
In this note, we consider this problem from a different viewpoint viz.,
the recent description of electrons as Kerr-Newman type Black Holes in a
Quantum Mechanical context\cite{r6,r7}. This will be shown to lead to the
conclusion that at small length scales applicable in the
cases mentioned above, electrons show precisely such an
anyonic behaviour with the "one third" effect built in. The corroboration from the
Laughlin\cite{r8} analysis will be seen and implications in hadronic
physics will emerge.\\
In the Kerr-Newman approach referred to, it was seen that outside the
Compton wavelength, the potential is given by
\begin{equation}
\frac{ee'}{r} = A_0 \approx  2c \hbar (\frac{mc^2}{\hbar}) \int \eta^{\imath j}
\frac{T_{\imath j}}{r} d^3 x'\label{e1}
\end{equation}
where $e'$ is the test charge and $T_{\imath j}$ are the usual stress
energy tensors. We next observe that the usual three dimensionality of space,
as pointed out by Wheeler arises due to the doubleconnectivity of Fermions
which takes place outside the Compton wavelength (cf.refs.\cite{r6,r7} and
references therein). But as we approach the Compton wavelength, we encounter
predominantly negative energy components and the above doubleconnectivity
and therefore three dimensionality disappears: In other words we encounter
two dimensions or one dimension. Indeed such a conclusion has been drawn
alternatively at small scales\cite{r9}.\\
Using the fact that each of the $T_{\imath j}$ in (\ref{e1}) is given by
$\frac{1}{3}\epsilon$ where $\epsilon$ is the energy density, it follows
from (\ref{e1}) that the particle would have the charge $\frac{2}{3}e$
or $\frac{1}{3}e$ in two or one dimensions.\\
Further as we encounter predominantly the negative energy two spinor of
the Dirac four spinor near the Compton wavelength, with negative helicity,
not only does the doubleconnectivity that is spinorial behaviour begin
to breakdown, that is anyonic behaviour emerges, but also the particles
would display handedness. Indeed in these low dimensions the Dirac
equation does exhibit helicity.\\
All this can be corroborated from a more conventional viewpoint (cf.ref.\cite{r8}).
Indeed the magnetic length $a_o$ becomes the Compton wavelength, as the
frequency $\omega = \frac{mc^2}{\hbar}$ in this case. Further as Laughlin
concluded, in this approach the ground state is a fractionally charged
Quantum fluid of quasi electrons.\\
An interesting consequence for hadronic physics follows by using Laughlin's
expression for the total energy viz.,
$$-\frac{1}{2} \pi^{1/2} e^2/R$$
Equating this with $m_ec^2$, the electron energy and as $R \sim$ of the
Compton wavelength, $\frac{\hbar}{mc}$ and $e$ from the above shows up as
$\sim \frac{e}{3}$ we can deduce that $m \sim 10^3 m_e$ that is the
constituents of the Quantum fluid show up with quark masses, their
fractional charge and handedness.


\begin{thebibliography}{99}
\bibitem {r1} Wildoer, J.W.G., et al, Nature\underline{391}, 1998, p59.
\bibitem {r2} Odom Teri Wang, Huang Jin-Lin, Philip Kim and Charles M. Lieber,
Nature\underline{391},1998, p62-64.
\bibitem {r3} Yang, J., and Su, W.P., Mod.Phys.Lett.B, \underline{7}(2), 1993, pp57-70.
\bibitem {r4} Li, D.P., Mod.Phys.Lett.B, \underline{7}(16), 1993, pp1103-1110.
\bibitem {r5} Wilczek, F., "Fractional Statistics and Anyon Superconductivity",
World Scientific, Singapore, 1990.
\bibitem {r6} Sidharth, B.G., Int.J. of Mod.Phys.A
\underline{13}(15), 1998, pp599ff.
\bibitem {r7} Sidharth, B.G.,  Ind.J. Pure \& Appd.Phys., \underline{35} (7), 1997,
456-471.
\bibitem {r8} Laughlin, R.B., Phys.Rev.Lett, \underline{50}(18), 1983, pp1395-1398.
\bibitem {r9} Altaiski, M.V., and Sidharth, B.G., Int.J.Theo.Phys., \underline{34}(12),
1995, pp2343-2352 and references therein.
\end{thebibliography}
\end{document}